\documentclass[aps,twocolumn,superscriptaddress,showpacs,letter,nofootinbib]{revtex4-1} 
\usepackage[utf8]{inputenc}
\usepackage{graphicx}
\usepackage{dcolumn}
\usepackage{bm}
\usepackage{cancel}
\usepackage{color}             
\usepackage{epsfig}
\usepackage{amssymb}
\usepackage{amsmath}
\usepackage{multirow}
\usepackage{hyperref}
\usepackage{cleveref}
\usepackage{lineno,color}
\usepackage{epstopdf}

\begin{document}

\title{Odderon contribution in light of the LHC low-$t$ data}

\author{E.~G.~S.~Luna}
\email{luna@if.ufrgs.br}
\affiliation{Instituto de F\'isica, Universidade Federal do Rio Grande do Sul, Caixa Postal 15051, 91501-970, Porto Alegre, Rio Grande do Sul, Brazil}
\author{M.~G.~Ryskin}
\email{ryskin@thd.pnpi.spb.ru}
\affiliation{Petersburg Nuclear Physics Institute, NRC Kurchatov Institute, \\Gatchina, St.~Petersburg, 188300, Russia}
\author{V.~A.~Khoze}
\email{v.a.khoze@durham.ac.uk}
\affiliation{Institute for Particle Physics Phenomenology, University of Durham, \\ Durham, DH1 3LE, UK}
 

\begin{abstract}

We perform the analysis of elastic scattering $pp$ and $\bar{p}p$ data at low momentum transfer $|t|<0.1$ GeV$^2$ within large collider energy interval $\sqrt{s}=50$ GeV - 13 TeV in order to evaluate quantitatively the possible Odderon contribution. We use the two-channel eikonal model, which naturally accounts for the screening of the Odderon amplitude by $C$-even (Pomeron) exchanges.

\end{abstract}


\maketitle

\section{Introduction}

The TOTEM publication ~\cite{Tot-13} of the measurements of the total cross section and the real part of the forward elastic $pp$ amplitude at 13 TeV, prompted a renewal of interest in the potential existence of the high-energy $C$-odd (Odderon) contribution.
This is because the observed value of the ratio of real to imaginary parts of the forward scattering amplitude, namely $\rho=(0.09-0.10)\pm 0.01$,  turned out to be noticeably smaller than the predicted value ($\rho=0.13-0.14$~\cite{Compete}) based on dispersion relations for the case of pure $C$-even interactions.

The new ATLAS/ALFA data recently confirmed this value of $\rho$~\cite{LHC1c}. However, the value of the total cross section at 13 TeV reported by the ATLAS/ALFA team, $\sigma_{tot}=104.68\pm1.09$ mb, is approximately 5\% lower than the average of values determined by TOTEM ($\sigma_{tot}=111.6\pm3.4$ mb, $\sigma_{tot}=109.5\pm3.4$ mb, and $\sigma_{tot}=110.3\pm 3.5$ mb)\footnote{Recall that the ATLAS/ALFA measurements at
$\sqrt{s} =$ 7 and 8 TeV \cite{LHC1a,LHC1b} are also
systematically lower than the corresponding TOTEM results \cite{LHC2, Sigt}.}, indicating that a smaller value of the real part of the $C$-even amplitude should be expected from the dispersion relations. 

The relatively small value of $\rho$ can be explained by the admixture of the $C$-odd amplitude, which survives at high LHC energies. Such amplitude with the intercept $\alpha_{Odd}$ close to 1 was predicted by the perturbative QCD~\cite{BKP, BLV} (see also the reviews~\cite{BE})  where at the lowest $\alpha_s$ order it is provided by the three gluon exchange diagrams. Alternatively, it was shown in~\cite{DL, BFLP} that if we do not assume a constant value for the real part of the nuclear amplitude near $|t|=0$, the whole ensemble of the elastic scattering low $|t|$ data (from which the total cross sections and the $\rho$-parameter for the forward amplitude were extracted) can be satisfactorily described without any Odderon contribution.\footnote{Note however that the analysis of \cite{DL} did not account for the Bethe phase in Coulomb-nuclear interference.}

Another indication in favor of the Odderon was claimed in~\cite{T-D0}, where the $\bar{p}p$ differential cross section $d\sigma/dt$ at $\sqrt s=1.96$ TeV was compared with the corresponding $pp$ cross section (measured at 2.76 TeV but extrapolated to 1.96 TeV) in the diffractive dip region. A clear difference was observed.

The problem, however, is that using this method, any inaccuracy in the extrapolation from one energy to another is treated as the additional (Odderon) contribution. In particular, we recall that using another prescription for this (from 2.76 to 1.96 TeV) extrapolation, the authors of~\cite{CNP} obtained for the Odderon signal significance less than 0.5 $\sigma$ (see Table 3 of~\cite{CNP}) analyzing the {\em same} data points.

Next, it should be noted that at very low $t$ close to zero and in the diffractive dip region, we deal with different $C$-odd contributions. To get a well-pronounced dip-bump structure near the dip in $pp$ scattering at 2.76 TeV and rather flat behavior of $\bar{p}p$ at 1.96 TeV, the real part of the Odderon $pp$ amplitude should be positive (in agreement with perturbative QCD expectation for the three gluon exchange case~\cite{DL}). On the other hand, to explain the low value of $\rho$ at $t=0$, we need a negative Odderon real part. This negative real part could be induced by non-perturbative effects.

In the present paper, we describe the full ensemble of the $pp$ and $\bar{p}p$ elastic differential cross-section data at low $|t|<0.1$ GeV$^2$ from $\sqrt s=50$ GeV up to 13 TeV. We use a two-channel eikonal model, which now includes the Odderon exchange amplitude. This allows us to quantitatively evaluate the size of the Odderon contribution and the parameters (the slope and the intercept) of the soft Pomeron trajectory.

In our analysis, the Odderon amplitude is not written separately but included in the eikonal. This inclusion allows the eikonal model to immediately account for the screening of the Odderon exchange by the Pomeron(s).

Moreover, we explore the tension between the $d\sigma/dt$ data measured by the TOTEM and ATLAS/ALFA Collaborations since the discrepancies in the results presented by the two experiments lead us to different scenarios for the behavior of the forward scattering amplitude and, consequently, for the parameters of both the Pomeron and Odderon.
 
The outline of this paper is as follows. In Sec. II we describe the model and collect the formulae used in our data description. The results are presented and discussed in Sec. III. In Sec. IV we draw our conclusions.

\section{Formalism}

Our analysis is focused on differential cross-section data involving very small values of $t$, which requires considering the Coulomb-nuclear interference (CNI) region. The full scattering amplitude, including the electromagnetic (Coulomb, $C$) and hadronic (Nuclear, $N$) interactions, can be expressed as
\begin{eqnarray}
 F^{C+N} = F^{N} + e^{i\alpha \phi(t)}F^{C} ,
\label{interf1}  
\end{eqnarray}
where the phase factor $\alpha \phi(t)$ describes the distortion of the pure amplitudes $F^{C}$ and $F^{N}$ arising from the simultaneous presence of both Coulombic and hadronic scattering. We adopt an expression for the  phase $\phi(t)$ derived from an eikonal approach \cite{Cahn01}, given by
\begin{eqnarray}
\phi (t) &=& \kappa \left[ \gamma + \ln \left( \frac{B |t|}{2} \right) + \ln \left( 1 + \frac{8}{B \Lambda^{2}} \right)  \right. \nonumber \\
& & + \left. \frac{4|t|}{\Lambda^{2}} \ln \left( \frac{\Lambda^{2}}{4|t|} \right) - \frac{2|t|}{\Lambda^{2}} \right] ,
\label{origcahn}
\end{eqnarray}
where $\kappa$ flips sign when going from $pp$ ($\kappa = -1$) to $\bar{p}p$ ($\kappa = +1$). In (\ref{origcahn}), $\Lambda^{2}$ is fixed at 0.71 GeV$^{2}$ (as determined in the dipole fit to the proton's electromagnetic form factor), and $B$ is the $t$ slope of elastic $d\sigma/dt\propto exp(Bt)$ cross-section. The 
 $\alpha$ in (\ref{interf1}) is the electromagnetic coupling, which is usually kept constant at the value of the fine structure constant, $\alpha(0)=1/137.03599911$ \cite{pdg}. Nevertheless, in cases where the Coulomb peak is used in the normalization of diffractive data, the contribution of vacuum polarization to the Coulomb force can be significant \cite{Cahn01,yennie01}. This effect results in the replacement of the fine structure constant with an electromagnetic coupling dependent on $q^{2}$,
\begin{eqnarray}
\alpha(q^2) = \frac{\alpha(0)}{1 - \frac{\alpha(0)}{3\pi } \ln \left( \frac{q^2 + m_{e}^{2}}{m_{e}^{2}} \right)} ,
\label{coupl2}
\end{eqnarray}
where $q^{2}=-t$ and $m_{e} = 0.510998918$ MeV \cite{pdg} is the mass of the electron. This corresponds to a one percent correction for $q^2=0.1$ GeV$^{2}$.

The Coulomb amplitude can be expressed as
\begin{eqnarray}
F^{C} = \kappa s\frac{2 \alpha }{|t|} G^2(t) .
\end{eqnarray}
Here, $G(t)$ is the electromagnetic form factor of the proton, described by the dipole form
\begin{eqnarray}
G(t) = \left[ \frac{\Lambda^{2}}{\Lambda^{2} + q^{2}}  \right]^{2} .
\end{eqnarray}

%

To account for the eikonalization, it is convenient to calculate the nuclear amplitude in terms of opacities.
The opacity function $\Omega_{i}(s,b)$ is related to the bare nuclear amplitude $F^{N}_{i}(s,t)$ through the Fourier-Bessel transform
\begin{eqnarray}
\Omega_{i} (s,b) = \frac{2}{s}\int^{\infty}_{0} q\, dq\, J_{0}(bq)\, F^{N}_{i}(s,t) ,
\label{opacity001}
\end{eqnarray}
where $i = {\Bbb P}, {\Bbb O}$ represent the Pomeron and Odderon exchanges, respectively.

The single Pomeron contribution is given by
\begin{eqnarray}
F^{N}_{\Bbb P}(s,t) = \beta_{\Bbb P}^{2}(t)\, \eta_{\Bbb P}(t) \left( \frac{s}{s_{0}} \right)^{\alpha_{\Bbb P}(t)} ,
\end{eqnarray}
where $\eta_{\Bbb P}(t) = -e^{-i\frac{\pi}{2}\alpha_{\Bbb P}(t)}$ is the even signature factor, $\beta_{\Bbb P}(t)$ is the elastic proton-Pomeron vertex, and
\begin{eqnarray}
\alpha_{\Bbb P}(t) = 1 + \epsilon + \alpha^{\prime}_{\Bbb P} t + \frac{m_{\pi}^{2}\beta_{\pi}^{2}}{32\pi^{3}}\, h(\tau)
\end{eqnarray}
is the Pomeron trajectory, where 
\begin{eqnarray}
h (\tau) &=& -\frac{4}{\tau}\, F_{\pi}^{2}(t) \left[  2\tau - (1+\tau)^{3/2} \ln \left( \frac{\sqrt{1+\tau}+1}{\sqrt{1+\tau}-1} \right) \right. \nonumber \\
 & & \left. + \ln \left( \frac{m^{2}}{m_{\pi}^{2}} \right) \right]  ,
\end{eqnarray}
with $\tau = 4m_{\pi}^{2}/|t|$, $F_{\pi}(t)=1/(1-t/m^{2}_{\rho})$, $m_{\pi}=139.6$ MeV, $m=1$ GeV, $m_{\rho}=0.776$ GeV, and $\beta_{\pi}/\beta_{I\!\!P}(0)=2/3$. The function $h$ accounts for the inclusion of the pion loop into the Pomeron trajectory~\cite{AG}.

The Odderon contribution is given by
\begin{eqnarray}
F^{N}_{\Bbb O}(s,t) =  \beta_{\Bbb O}^{2}(t)\, \eta_{\Bbb O}(t) \left( \frac{s}{s_{0}} \right)^{\alpha_{\Bbb O}(t)} ,
\label{odd1}
\end{eqnarray}
where $\eta_{\Bbb O}(t) = -ie^{-i\frac{\pi}{2}\alpha_{\Bbb O}(t)}$ is the odd signature factor, $\beta_{\Bbb O}(t)$ is the elastic proton-Odderon vertex,
and $\alpha_{\Bbb O}(t) = 1$ is the Odderon trajectory.

The $t$-dependence of the $\beta$ vertices is parameterized accounting for the observed deviation from a pure exponential behavior of the low-$|t|$ $d\sigma/dt$ data at LHC energies, as identified by the TOTEM Collaboration \cite{Tot-1,Tot-2,Tot-3}. To get a better fit in the small-$t$ region, the TOTEM group has extended the pure exponential to a cumulant expansion,
\begin{eqnarray}
\frac{d\sigma}{dt}(t) = \left. \frac{d\sigma}{dt} \right|_{t=0} \exp \left( \sum_{n=1}^{N_{b}} b_{n} t^{n}  \right) ,
\end{eqnarray}
where the optimal fit was achieved for $N_{b}=3$, yielding $\chi^{2}/DoF=1.22$ and a corresponding $p$-value of 8.0\%. The same was done by the ATLAS/ALFA group

Based on this result, we have written the Pomeron- and Odderon-proton vertices as 
\begin{eqnarray}
\beta_{\Bbb P}(t) = \beta_{\Bbb P}(0) e^{\left( At+Bt^{2}+Ct^{3} \right)/2}
\label{bP}
\end{eqnarray}
and
\begin{eqnarray}
\beta_{\Bbb O}(t) = \beta_{\Bbb O}(0) e^{Dt/2} ,
\label{Od-sl}  
\end{eqnarray}
respectively.

In order to allow for the low mass diffractive dissociation, the Good-Walker formalism \cite{GW} is used, which provides a convenient form to incorporate $p \to N^{*}$ diffractive dissociation. This approach introduces diffractive eigenstates $|\phi_{k}\rangle$ that diagonalize the interaction $T$ matrix. In our two-channel model ($k=1, 2$), the proton wave function is described by two components with equal weights, namely $|p\rangle =\sqrt{\frac{1}{2}} \left( |\phi_1\rangle +|\phi_2\rangle \right)$. To minimize the number of free parameters, we take the same $t$ dependence for both components. The Pomeron and Odderon couplings to the two diffractive states $|\phi_{k}\rangle$ are
\begin{eqnarray}
\beta_{i,k}(t) = (1\pm \gamma)\beta_{i}(t),  
\end{eqnarray}
with $i={\Bbb P}$ or ${\Bbb O}$, and $\gamma=0.55$.\footnote{With $\gamma=0.55$ at $\sqrt s=7$ TeV we get the low mass dissociation cross-section $\sigma^D=4.7$ mb in agreement with the TOTEM result~\cite{sigD} $\sigma^D=2.6\pm 2.2$ mb.}

The eikonalized amplitude in $(s,t)$-space is then given by \cite{kmr2000,kmr2001}
\begin{eqnarray}
{\cal A}(s,t) &=& is \int^{\infty}_{0} b\, db\, J_{0}(bq) \left[ 1 -\frac{1}{4}\, e^{i(1+\gamma)^{2}\Omega(s,b)/2} \right. \nonumber \\
& & \left. -\frac{1}{2}\, e^{i(1-\gamma^{2})\Omega(s,b)/2} - \frac{1}{4}\, e^{i(1-\gamma)^{2}\Omega(s,b)/2} \right], \nonumber \\
\label{doubeik001}
\end{eqnarray}
where $\Omega(s,b)$ is the total opacity.

We consider two versions for the total opacity with different signs for the Odderon contribution. In the first version, referred to as `Model {\bf I}', we have
\begin{eqnarray}
\Omega(s,b) = \Omega_{\Bbb P}(s,b)  \mp \Omega_{\Bbb O} (s,b);
\end{eqnarray}
in the second version, called `Model {\bf II}', we have
\begin{eqnarray}
\Omega(s,b) = \Omega_{\Bbb P}(s,b)  \pm \Omega_{\Bbb O}(s,b) ;
\end{eqnarray}
in both cases the upper sign is for $pp$ and the lower sign is for $\bar{p}p$.\footnote{Recall that the sign of the Odderon exchange amplitude is not known from the beginning. In our parameterization (10) we keep $\beta_{\Bbb O}^{2}>0$ to be positive. That is, model II corresponds to the case when the Odderon contribution to the real part of $pp$ amplitude is negative. This is opposite to the sign expected in perturbative QCD for the leading order $C$-odd three gluon exchange diagram \cite{DL,fukugita1}. However, at $t\to 0$ the situation is more complicated and may not be described by the pure three gluon exchange. In particular, within the quark-diquark model of the proton, the Odderon coupling $\beta^2_{\Bbb O}$ nullifies at $t=0$ (for the point-like diquark) and becomes positive at larger $|t|$ values. In such a case at $t\to 0$ the major $C$-odd contribution comes from the Pomeron-Odderon cut and the real part of $C$-odd $pp$ amplitude becomes negative.}

The total cross section and the $\rho$ parameter are expressed in terms of the nuclear eikonalized amplitude ${\cal A}(s,t)$:
\begin{eqnarray}
\sigma_{tot}(s) = \frac{4 \pi}{s} \textnormal{Im}\, {\cal A}(s,t=0),
\label{sigtot}
\end{eqnarray}
\begin{eqnarray}
\rho(s) = \frac{\textnormal{Re}\, {\cal A}(s,t=0)}{\textnormal{Im}\, {\cal A}(s,t=0)} .
\label{rhopar}
\end{eqnarray}

Considering the eikonalized nuclear amplitude (\ref{doubeik001}) in the presence of electromagnetic and hadronic interactions, our full scattering amplitude will finally be written as
\begin{eqnarray}
 F^{C+N}(s,t) = {\cal A}(s,t) + e^{i\alpha \phi(t)}F^{C}(s,t) .
\label{interf2}  
\end{eqnarray}

Thus we can write the differential and the total elastic cross sections as
\begin{eqnarray}
\label{th}
\frac{d\sigma}{dt}(s,t) = \frac{\pi}{s^{2}}\, \left| {\cal A}(s,t) + e^{i\alpha \phi}{\cal F}^{C}(s,t) \right|^{2} ,
\end{eqnarray}
\begin{eqnarray}
\label{th2}
\sigma_{el}(s) = \frac{\pi}{s^{2}}\, \int_{-\infty}^{0} dt  \left| {\cal A}(s,t)  \right|^{2} .
\end{eqnarray}

\section{Results}

The LHC has provided highly precise measurements of diffractive processes, allowing stringent constraints on the scattering amplitude behavior at high energies. These measurements, particularly of the $\rho$ parameter and the total and differential cross sections from ATLAS/ALFA and TOTEM experiments play a crucial role
 in the accurate determination of the Pomeron and Odderon parameters.
However, while the measurements of the $\rho$ parameter at $\sqrt{s}=13$ TeV by both Collaborations are consistent \cite{Tot-13, LHC1c}, the total and differential cross sections at $\sqrt{s}=7$, 8, and 13 TeV reveal some tension between the TOTEM \cite{Tot-1, Tot-13,totem7a, Sigt, Sigt2} and ATLAS/ALFA \cite{LHC1a, LHC1b, LHC1c} results. This data discrepancy suggests different scenarios for the rise of the total cross section and, hence, the parameters of the Pomeron and the Odderon \cite{luna1,petrov1}.

In order to systematically explore the tension between the TOTEM and ATLAS/ALFA results, we perform global fits to the $pp$ and $\bar{p}p$ differential cross sections considering three distinct data sets: one comprising solely the TOTEM measurements, another consisting solely of the ATLAS/ALFA results, and the third one combining both. These data sets are complemented by the $d\sigma/dt$ data spanning from CERN-ISR, $S\bar{p}pS$, to Tevatron energies.
Specifically, the three data ensembles can be defined as follows: \\

{\bf Ensemble A}: $\left. \frac{d\sigma^{\bar{p}p,pp}}{dt} \right|_{\footnotesize \textnormal{CERN-ISR}}$ + $\left. \frac{d\sigma^{\bar{p}p}}{dt} \right|_{\footnotesize S\bar{p}pS}$ + $\left. \frac{d\sigma^{\bar{p}p}}{dt} \right|_{\footnotesize \textnormal{Tevatron}}$ + $\left. \frac{d\sigma^{pp}}{dt} \right|_{\footnotesize \textnormal{ATLAS/ALFA}}$ \\

{\bf Ensemble T}: $\left. \frac{d\sigma^{\bar{p}p,pp}}{dt} \right|_{\footnotesize \textnormal{CERN-ISR}}$ + $\left. \frac{d\sigma^{\bar{p}p}}{dt} \right|_{\footnotesize S\bar{p}pS}$ + $\left. \frac{d\sigma^{\bar{p}p}}{dt} \right|_{\footnotesize \textnormal{Tevatron}}$ + $\left. \frac{d\sigma^{pp}}{dt} \right|_{\footnotesize \textnormal{TOTEM}}$ \\

{\bf Ensemble A$\oplus$T}: $\left. \frac{d\sigma^{\bar{p}p,pp}}{dt} \right|_{\footnotesize \textnormal{CERN-ISR}}$ + $\left. \frac{d\sigma^{\bar{p}p}}{dt} \right|_{\footnotesize S\bar{p}pS}$ + $\left. \frac{d\sigma^{\bar{p}p}}{dt} \right|_{\footnotesize \textnormal{Tevatron}}$ + $\left. \frac{d\sigma^{pp}}{dt} \right|_{\footnotesize \textnormal{ATLAS/ALFA}}$ + $\left. \frac{d\sigma^{pp}}{dt} \right|_{\footnotesize \textnormal{TOTEM}}$ \\

The $\left. d\sigma^{\bar{p}p,pp}/dt \right|_{\footnotesize \textnormal{CERN-ISR}}$ data enclose measurements of the differential cross section for $pp$ scattering at $\sqrt{s} =$ 52.8 \cite{exp_amos1}, and 62.5 \cite{exp_amaldi1} GeV and $\bar{p}p$ scattering  at $\sqrt{s} =$ 52.6 \cite{exp_amos1}, 53 \cite{exp_breakstone1}, 62 \cite{exp_breakstone1}, and 62.3 GeV \cite{exp_amos1}. The $\left. d\sigma^{\bar{p}p}/dt \right|_{\footnotesize S\bar{p}pS}$ data set comprise differential cross section data for $\bar{p}p$ channel at $\sqrt{s}=540$ \cite{exp_arnison1} and 546 GeV \cite{exp_bernard1}.
The $\left. d\sigma^{\bar{p}p}/dt \right|_{\footnotesize \textnormal{Tevatron}}$ data set consist of  $d\sigma/dt$ data for the $\bar{p}p$ channel at $\sqrt{s}=1800$ GeV \cite{Tev1,Tev2}. The $\left. \frac{d\sigma^{pp}}{dt} \right|_{\footnotesize \textnormal{ATLAS/ALFA}}$ contains measurements of the differential cross section for $pp$ scattering at $\sqrt{s}=$ 7 \cite{LHC1a}, 8 \cite{LHC1b}, and 13 TeV \cite{LHC1c} obtained using the ATLAS Roman Pot system ALFA \cite{romanalfa}.
Finally, the $\left. \frac{d\sigma^{pp}}{dt} \right|_{\footnotesize \textnormal{TOTEM}}$ data set comprise differential cross section data for $pp$ channel at $\sqrt{s}=$ 7 \cite{totem7a}, 8 \cite{Tot-1}, and 13 TeV \cite{Tot-13} measured by the TOTEM experiment.

Hence, only the elastic scattering data are included in the analysis since the experimental $\sigma_{tot}$ and $\rho$ values were obtained from fitting the same $d\sigma/dt$ points already included in our analysis. Of course, to be sensitive to the $\rho$, we keep all the very small $|t|$ data points and account for the CNI region. Moreover, to minimize the number of free parameters, we start from the relatively large energy, namely $\sqrt s > 50$ GeV, where the secondary Reggeon contribution can be neglected, and fix the Odderon trajectory ($\alpha_{\Bbb O}(t)=1$).

The values of $B$ to be used in the Coulomb phase (equation (\ref{origcahn})) are the ones obtained by determining the differential cross sections at different center-of-mass energies, as indicated in the original articles. Specifically, the values are 12.86 GeV$^{-2}$ for the $pp$ at $\sqrt{s}=$ 52.8 GeV, 13.21 GeV$^{-2}$ for the the $pp$ at $\sqrt{s}=$ 62.5 GeV,
13.36 GeV$^{-2}$ for the $p\bar{p}$ at $\sqrt{s}=$ 52.6 GeV,
11.5 GeV$^{-2}$ for the $p\bar{p}$ at $\sqrt{s}=$ 53 GeV,
11.12 GeV$^{-2}$ for the $p\bar{p}$ at $\sqrt{s}=$ 62 GeV,
13.47 GeV$^{-2}$ for the $p\bar{p}$ at $\sqrt{s}=$ 62.3 GeV,
17.1 GeV$^{-2}$ for the $p\bar{p}$ at $\sqrt{s}=$ 540 GeV,
15.3 GeV$^{-2}$ for the $p\bar{p}$ at $\sqrt{s}=$ 546 GeV,
16.3 GeV$^{-2}$ for the $p\bar{p}$ at $\sqrt{s}=$ 1800 GeV (E710),
16.98 GeV$^{-2}$ for the $p\bar{p}$ at $\sqrt{s}=$ 1800 GeV (CDF),
19.73 GeV$^{-2}$ for the $pp$ at $\sqrt{s}=$ 7 TeV (ATLAS/ALFA), 
19.89 GeV$^{-2}$ for the $pp$ at $\sqrt{s}=$ 7 TeV (TOTEM),
19.74 GeV$^{-2}$ for the $pp$ at $\sqrt{s}=$ 8 TeV (ATLAS/ALFA), 
19.90 GeV$^{-2}$ for the $pp$ at $\sqrt{s}=$ 8 TeV (TOTEM),
21.14 GeV$^{-2}$ for the $pp$ at $\sqrt{s}=$ 13 TeV (ATLAS/ALFA), and
20.40 GeV$^{-2}$ for the $pp$ at $\sqrt{s}=$ 13 TeV (TOTEM).

Once our ensembles are defined, we start carrying out a fit to the Ensemble A$\oplus$T. In our analyses, we fit the CERN-ISR data with $|t|\leq 0.2$ GeV$^{2}$ and to the $S\bar{p}pS$, Tevatron, and LHC data with $|t|\leq 0.1$ GeV$^{2}$. We use a $\chi^{2}$ fitting procedure, with the value of $\chi^{2}_{min}$ distributed as a $\chi^{2}$ distribution with $\nu$ degrees of freedom. We adopt an interval $\chi^{2}-\chi^{2}_{min}$ corresponding to a 90\% confidence level (CL).

Since the absolute values of cross sections measured at the same energy by different groups do not agree, we have introduced normalization factors $N_{i}$ for high-energy $d\sigma/dt$ data, with $i=7[A]$, $8[A]$, and $13[A]$ for the ATLAS/ALFA data and $i=7[T]$, $8[T]$, and $13[T]$ for the TOTEM data. Here, the numbers within the indices $i$ correspond to the values of $\sqrt{s}$, namely 7, 8, and 13 TeV. Analogous normalization factors are introduced for the Tevatron data with $i=1.8[E]$ and $i=1.8[C]$, i.e. $N_{1.8[E]}$ for the E710 data and $N_{1.8[C]}$ for the CDF data. Despite being the only data set measured at $\sqrt s = $ 546 GeV, we also included a normalization factor for $\left. d\sigma^{\bar{p}p}/dt \right|_{\sqrt{s}=546\, \textnormal{GeV}}$, namely $N_{546}$.
Furthermore, when dealing with the data sets incorporating normalization factors $N_{i}$, we make use of the formula
\begin{equation}
\label{chi}
\chi^2=\sum_{ij}\frac{(N_ids^{th}_{ij}-ds^{exp}_{ij})^2}{(\delta^{rem}_{ij})^2} + \sum_i \frac{(1-N_i)^2}{\delta^2_i} 
\end{equation}
to calculate the total $\chi^2$ value, where $i$, as already specified, denotes the particular set of data while $j$ denotes the point $t_j$ in this set of data; $ds^{th}_{ij}$ is the theoretically calculated $d\sigma/dt$ cross section (\ref{th}) while $ds^{exp}_{ij}$ is the value measured at the same $ij$ point experimentally; $\delta_i$ is the normalization uncertainty of the given ($i$) set of data and $\delta^{rem}_{ij}$ is the remaining error at the point $ij$ calculated as $(\delta^{rem}_{ij})^2=\delta^2_{tot,ij}-(\delta_i\cdot ds^{exp}_{ij})^2$. As a rule the value of $\delta^{rem}$ is dominantly the statistical error.\footnote{A similar approach was used in \cite{Sel}.}

The values of the free parameters determined by the fit to the Ensemble A$\oplus$T, as derived from Models I and II, are listed in Table~\ref{tab001}. These results were obtained by fixing in (\ref{Od-sl}) the Odderon amplitude slope $D=A/2$.
The second and third columns exhibit the outcomes obtained by permitting the normalization factors to fluctuate within the interval [0.85,  1.15]; the results of these fits are shown in Figs. 1-6. We first observe that the parameters related to the Pomeron are not sensitive to the chosen model (Model I or II) since their values are compatible with each other considering the associated uncertainties. The same is not observed concerning the Odderon coupling $\beta_{\Bbb O}(0)$: its value is consistent with zero (error significantly surpassing central value) in the case of Model I. Consequently, it is clear that a positive Odderon contribution to the real part of $pp$ elastic amplitude is rejected by the data, resembling a scenario where the Pomeron dominates the scattering amplitude. Therefore, we consider the Model I as the model {\em without} the Odderon.
Moreover, from the statistical standpoint, the fit using Model II is appreciably better than that of Model I, as the resultant $\chi^{2}/\nu$ values are 1.11 and 1.44, respectively. Hence the inclusion of a negative Odderon contribution to the real part of $pp$ elastic amplitude decreases the value of $\chi^{2}$ when compared with the value obtained in the analysis using the model without the Odderon; specifically, the decrease is around 25\%.

The value of $N_{13[T]}$ each time is at the edge of the allowed interval, indicating some problem with the TOTEM 13 TeV data. All other $N_i$ factors are {\em inside} the interval. For a larger interval, we get a smaller $\epsilon$ (i.e., cross section grows slower with energy) and correspondingly a smaller Odderon coupling $\beta_{\Bbb O}(0)$ since a smaller $\epsilon$ leads to a smaller $\rho$ for a pure even contribution. With a larger allowed interval, we get a smaller $\chi^{2}$ (at least up to [0.6-1.4] case); however, we would not consider so large normalization factor as a realistic value. Therefore, we chose the [0.85-1.15] interval as the main one, with the fourth column of Table~\ref{tab001} (for the interval [0.80,1.20]) provided merely to illustrate the trend.

The effect of incorporating the Odderon becomes notably significant when analyzing specific subsets of data. In particular, it becomes more evident in part of the data in the ISR region, notably in data with $0.1 \lesssim |t| \lesssim 0.2$ GeV$^{2}$ at energies $\sqrt{s} = $ 53 and 62 GeV (Fig. 1). Although less pronounced, its influence is also visible in part of the data in the $S\bar{p}pS$ region, namely the data with $0.05 \lesssim |t| \lesssim 0.1$ GeV$^{2}$ at energy $\sqrt{s} = $ 540 GeV (Fig. 2); in the data in the Tevatron region, specifically in the data with $0.03 \lesssim |t| \lesssim 0.1$ GeV$^{2}$ at energy $\sqrt{s} = $ 1.8 TeV (Fig. 2); and in the LHC data, particularly in the CNI region at the energies $\sqrt{s} = $ 7, 8, and 13 TeV (Figs. 3 and 4).

The influence of the Odderon is also apparent in our predictions for the behavior of the total cross section $\sigma_{tot}(s)$, the total elastic cross section $\sigma_{el}(s)$ and the parameter $\rho (s)$. After introducing the Odderon, we can see in Fig. 5 a slight difference between the $pp$ (solid curve) and the $\bar{p}p$ (dashed curve) channels, both in the case of $\sigma_{tot}$ and $\sigma_{el}$. For Model I, the $pp$ (dotted curve) and $\bar{p}p$ (dashed-dotted curve) channels are indistinguishable since then the scattering amplitude is dominated asymptotically only by the even terms, the total cross-section difference behaves as $\left| \Delta \sigma \right| = \left| \sigma_{tot}^{\bar{p}p} - \sigma_{tot}^{pp} \right| \to$ 0 in the limit $s\to \infty$.  

The influence of the Odderon also becomes particularly evident when examining the Model II predictions for the $\rho$ parameter (Fig. 6): there is a clear separation between the $pp$ and $\bar{p}p$ channels. As illustrated in Table~\ref{tab001}, the model predictions for $\rho^{pp}$ and $\rho^{\bar{p}p}$ at $\sqrt{s} = 13$ TeV are, respectively, 0.111 and 0.119. Conversely, under the Model I, the channels $pp$ and $\bar{p}p$ predictions are identical and equal to 0.114. Table~\ref{tab002} shows the predictions for the high energy total cross section and $\rho$ parameter, obtained using Model I and Model II. Note that similar values of $\sigma_{tot}=107.6\pm 1.7$ mb and $\rho=0.11\pm 0.01$ at 13 TeV were derived in \cite{PT} fitting the low $|t|$ TOTEM 13 TeV data {\em only} but introducing the normalization coefficient $\lambda$ (analogous to $1/N_{13,T}$ in our Table~\ref{tab001}). In this case, the fit was normalized to the Coulomb scattering.

Regarding the behavior of the Pomeron- and Odderon-proton vertices, note that the parameters $A$ and $D$ are not independent, as they must satisfy the unitarity constraint $D<A$. In this way, as previously indicated, we choose $D=A/2$ in our analysis. We have studied other relations between $A$ and $D$, specifically the cases $D=0.1A$, $D=0.3A$, $D=0.7A$, and $D=0.9A$. As illustrated in Table~\ref{tab003}, these choices do not affect the global fits and, therefore, the model predictions. While the Pomeron ($C$-even amplitude) parameters remain relatively stable with respect to variations in the Odderon slope $D$, the value of the Odderon coupling $\beta_{\Bbb O}(0)$ decreases monotonically as $D$ increases (see Table~\ref{tab003}). This can be explained by the fact that for a smaller $D$, the Odderon contribution is concentrated at a smaller impact parameter $b$, where screening by the $C$-even amplitude is stronger. Thus, we need larger $\beta_{\Bbb O}(0)$ to get the same final result.

We investigated the effect of replacing, in Ensemble A$\oplus$T, the data set of the differential cross section $\bar{p}p$ at $\sqrt{s} = 546$ GeV \cite{exp_bernard1} by the measurement of the differential $dN/dt$ distribution performed by the UA4/2 Collaboration at $\sqrt{s} =$ 541 GeV \cite{UA42} (see Fig. 7). This study was to verify whether using the $dN/dt$ distribution, carried out with a substantial reduction in statistical uncertainty and more rigorous control of systematic effects than the previous measurement at $\sqrt{s} = $ 546 GeV, could affect our results and predictions for the LHC energy. Our analysis revealed that using the distribution $dN/dt$ instead of $d\sigma/dt$ in Ensemble A$\oplus$T does not affect the results already presented in this paper.

We further conducted an analysis focusing only on the ATLAS/ALFA or TOTEM data. As depicted in Table~\ref{tab003}, the ATLAS/ALFA results show slightly better agreement with their respective  Ensemble. However, accounting for all normalization factors $N_i$, the value $\chi^2/\nu=1.11$ remains highly satisfactory for the analysis using the Ensemble A$\oplus$T.\footnote{A slightly larger $\chi^2/\nu$ in the pure TOTEM case (Ensemble T) could be caused by the fact that in TOTEM data of 13 TeV, some oscillations occur in the $t$ behavior of $d\sigma^{pp}/dt$ (see e.g. \cite{Sel, Per}) that are not accounted for within our parameterization of the high-energy elastic amplitude. }

\section{Conclusions}

The differential proton-proton and proton-antiproton cross sections $d\sigma/dt$ at low $|t|<0.1$ GeV$^2$  and collider energies (from $\sqrt s>50$ GeV to 13 TeV) are successfully described ($\chi^2/\nu=1.11$) within the two-channel eikonal
model. To avoid the double counting we do not include in the fit the $\sigma_{tot}$ and $\rho$ data (which were obtained from the description of the same $d\sigma/dt$ data points) but rather extend our fit to the very low $|t|$ values
describing the Coulomb-nuclear interference region. The model accounts for the screening of the Odderon contribution by the Pomerons including the $C$-even (Pomeron) and $C$-odd (Odderon) multiple exchanges. That is the opacity, $\Omega$, is written as the sum of the bare Pomeron and the bare Odderon amplitudes.

To resolve the discrepancy between the TOTEM and ATLAS/ALFA (CDF and E710 in the Tevatron case) data we introduce the normalization coefficients, $N_{i}$ writing the theoretical prediction as $d\sigma^{exp}/dt=N_id\sigma^{Th}/dt$. The global analysis chooses $N_{i}>1$ for TOTEM and $N_{i}>1$ for ATLAS/ALFA. The deviation of the values of $N_i$ from 1 compared with the published luminosity uncertainties is included in the total $\chi^2$.\\

We show that the presence of $C$-odd (Odderon) contribution essentially improves the fit (see Table~\ref{tab001}) however it does not noticeably change the predicted value of $\rho^{pp}$ at 13 TeV.
A larger Odderon amplitude is constrained by the $S\bar{p}pS$ data.

For completeness, we present also the results obtained by fitting only the TOTEM or only the ATLAS/ALFA data.

The main lessons about the Odderon coming from this study are:
\begin{itemize}
\item The description using the Odderon improves the fit (the $\chi^2/\nu$ is the lowest one).
\item The sign of the Odderon amplitude needed to describe the very low $|t|$ data is opposite to that predicted by the perturbative QCD three-gluon exchange contribution.
\item
 The quality of the description weakly depends on the Odderon $t$-slope, $D$, (leading to practically the same values of $\sigma_{tot}$ and $\rho$ ). However, for smaller $D$ we need larger coupling $\beta_{\Bbb O}$ to compensate for a stronger absorption caused by the Pomeron screening at small impact parameters $b$.
 \item The Odderon-proton coupling, $\beta_{\Bbb O}$, is smaller than that for the Pomeron, $\beta_{\Bbb P}$. For $D=A/2$ we get
  $\beta_{\Bbb O}/\beta_{\Bbb P}=0.40$, however after accounting for screening by the Pomeron the final $C$-odd 
  contribution to $\rho$ at 13 TeV becomes quite small,
$\delta\rho=(\rho^{\bar pp}-\rho^{pp})/2\leq 0.004$ (see Table~\ref{tab001}) and it will be challenging to enlarge it. Otherwise, we will get too large $\rho^{\bar pp}$ at $\sqrt s \sim 541$ GeV in disagreement with the data \cite{UA42}.

\end{itemize}

\section*{Acknowledgment}

This research was partially supported by the Conselho Nacional de Desenvolvimento Cient\'{\i}fico e Tecnol\'ogico under Grant No. 307189/2021-0.

\begin{table*}[h]
\centering
\caption{Values of the parameters obtained in the global fits to Ensemble $A\oplus T$. }
\begin{ruledtabular}
\begin{tabular}{ccccc}
 & {\bf \footnotesize Model I}  & {\bf \footnotesize Model II} & {\bf \footnotesize Model II}  \\
 \hline \\ [-0.3cm]
$\beta_{\Bbb P}(0)$ & 2.247$\pm$0.013 & 2.259$\pm$0.016 & 2.307$\pm$0.022 \\
$\epsilon$ & 0.1173$\pm$0.0021 & 0.1180$\pm$0.0020 & 0.1134$\pm$0.0019 \\
$\alpha^{\prime}_{I\!\!P}$ (GeV$^{-2}$) & 0.124$\pm$0.024 & 0.128$\pm$0.022 & 0.133$\pm$0.023 \\
$A$ (GeV$^{-2}$) & 5.01$\pm$0.20 & 4.78$\pm$0.21 & 4.72$\pm$0.21 \\
$B$ (GeV$^{-4}$) & 6.61$\pm$0.99 & 6.7$\pm$1.1 & 6.9$\pm$1.2 \\
$C$ (GeV$^{-6}$) & 20.4$\pm$5.7 & 17.7$\pm$4.0 & 17.0$\pm$4.2 \\
$\beta_{\Bbb O}(0)$ & (0.15 $\times 10^{-4}$) $\pm$39 & 0.90$\pm$0.18 & 0.88$\pm$0.18 \\
$N_{546}$ & 0.941 & 0.933 & 0.958 \\
$N_{1.8[E]}$ & 0.923 & 0.912 & 0.944 \\
$N_{1.8[C]}$ & 1.087 & 1.070 & 1.109 \\ 
$N_{7[A]}$ & 1.015 & 1.015 & 1.056 \\
$N_{8[A]}$ & 1.003 & 1.003 & 1.045 \\
$N_{13[A]}$ & 1.009 & 1.009 & 1.052 \\
$N_{7[T]}$ & 1.077 & 1.077 & 1.121 \\
$N_{8[T]}$ & 1.121 & 1.121 & 1.167 \\
$N_{13[T]}$ & 1.150 & 1.150 & 1.200 \\ [0.05cm]
\hline \\ [-0.3cm]
$\rho^{pp} (\sqrt{s}=13\  \textnormal{TeV})$ & 0.114 & 0.111 & 0.109 \\
$\rho^{\bar{p}p} (\sqrt{s}=13\  \textnormal{TeV})$ & 0.114 & 0.119 & 0.116 \\
\hline \\ [-0.3cm]
Allowed $N_{i}$ interval & [0.85,1.15] & [0.85,1.15] & [0.80,1.20] \\
$\nu$ & 504  & 504 & 504 \\
$\chi^{2}/\nu$ & 1.44  & 1.11 & 1.03 
\end{tabular}
\end{ruledtabular}
\label{tab001}
\end{table*}

\begin{table*}
\centering
\caption{Predictions for $\sigma_{tot}^{\bar{p}p,pp}$, $\sigma_{el}^{\bar{p}p,pp}$, and $\rho^{\bar{p}p,pp}$ using Models I and II. These results were derived for the scenario with $D=A/2$.}
\begin{ruledtabular}
\begin{tabular}{ccccccc}
 & \multicolumn{3}{c}{Model I} & \multicolumn{3}{c}{Model II}  \\
\cline{2-4} \cline{5-7} \\ [-0.2cm]
$\sqrt{s}$ (TeV) & $\sigma^{pp}_{tot}$ $|$ $\sigma^{\bar{p}p}_{tot}$ (mb) & $\sigma^{pp}_{el}$ $|$ $\sigma^{\bar{p}p}_{el}$ (mb) & $\rho^{pp}$ $|$ $\rho^{\bar{p}p}$ & $\sigma^{pp}_{tot}$ $|$ $\sigma^{\bar{p}p}_{tot}$ (mb) & $\sigma^{pp}_{el}$ $|$ $\sigma^{\bar{p}p}_{el}$ (mb) & $\rho^{pp}$ $|$ $\rho^{\bar{p}p}$ \\
0.541 & 64.2 $|$ 64.2 & 13.2 $|$ 13.2 & 0.130 $|$ 0.130 & 63.8 $|$ 64.1 & 13.3 $|$ 13.5 & 0.117 $|$ 0.144 \\
1.8 & 78.0 $|$ 78.0 & 17.6 $|$ 17.6 & 0.124 $|$ 0.124 & 77.6 $|$ 77.8 & 17.7 $|$ 17.9 & 0.116 $|$ 0.133 \\
7 & 95.9 $|$ 95.9 & 23.9 $|$ 23.9 & 0.117 $|$ 0.117 & 95.7 $|$ 95.9 & 24.0 $|$ 24.2 & 0.113 $|$ 0.123 \\
8 & 97.9 $|$ 97.9 & 24.5 $|$ 24.5 & 0.116 $|$ 0.116 & 97.6 $|$ 97.8 & 24.7 $|$ 24.8 & 0.113 $|$ 0.122 \\
13 & 105.1 $|$ 105.1 & 27.2 $|$ 27.2 & 0.114 $|$ 0.114 & 104.9 $|$ 105.1 & 27.3 $|$ 27.4 & 0.111 $|$ 0.119 \\
\end{tabular}
\end{ruledtabular}
\label{tab002}
\end{table*}

\begin{table*}
\centering
\caption{Results using Model II}
\begin{ruledtabular}
\begin{tabular}{cccccc}
 & & \multicolumn{2}{c}{Ensemble $A$} & & \\
$D$ (GeV$^{-2}$) & $0.1A$ & $0.3A$ & $0.5A$ & $0.7A$ & $0.9A$ \\
$\beta_{\Bbb O}(0)$ & 0.93$\pm$0.22 & 0.85$\pm$0.22 & 0.80$\pm$0.21 & 0.77$\pm$0.19 & 0.74$\pm$0.18 \\
$\beta_{\Bbb P}(0)$ & 2.370$\pm$0.035 & 2.384$\pm$0.036 & 2.386$\pm$0.037 & 2.386$\pm$0.040 & 2.388$\pm$0.039 \\
$\nu$ & 332 & 332 & 332 & 332 & 332 \\
 $\chi^{2}/\nu$ & 0.96 & 0.97 & 0.97 & 0.97 & 0.96 \\
\hline \\ [-0.3cm]
$\rho^{pp} (\sqrt{s}=13\  \textnormal{TeV})$ & 0.105 & 0.105 & 0.105 & 0.104 & 0.104 \\
$\rho^{\bar{p}p} (\sqrt{s}=13\  \textnormal{TeV})$ & 0.113 & 0.112 & 0.113 & 0.114 & 0.114 \\
$\sigma_{tot}^{pp} (\sqrt{s}=13\  \textnormal{TeV})$ (mb)& 98.0 & 98.0 & 98.0 & 98.0 & 98.0 \\
$\sigma_{tot}^{\bar{p}p} (\sqrt{s}=13\  \textnormal{TeV})$ (mb) & 98.2 & 98.2 & 98.2 & 98.2 & 98.1 \\  
\hline   
\hline \\ [-0.3cm]
 & & \multicolumn{2}{c}{Ensemble $T$} & & \\  
$D$ (GeV$^{-2}$) & $0.1A$ & $0.3A$ & $0.5A$ & $0.7A$ & $0.9A$ \\
$\beta_{\Bbb O}(0)$ & 1.09$\pm$0.22 & 0.96$\pm$0.18 & 0.90$\pm$0.16 & 0.86$\pm$0.15 & 0.83$\pm$0.14 \\
$\beta_{\Bbb P}(0)$ & 2.236$\pm$0.022 & 2.258$\pm$0.016 & 2.260$\pm$0.016 & 2.260$\pm$0.017 & 2.259$\pm$0.018 \\
$\nu$ & 418 & 418 & 418 & 418 & 418 \\
$\chi^{2}/\nu$ & 1.28 & 1.30 & 1.29 & 1.28 & 1.27 \\
\hline \\ [-0.3cm]
$\rho^{pp} (\sqrt{s}=13\  \textnormal{TeV})$ & 0.112 & 0.112 & 0.111 & 0.111 & 0.110 \\
$\rho^{\bar{p}p} (\sqrt{s}=13\  \textnormal{TeV})$ & 0.119 & 0.118 & 0.119 & 0.119 & 0.120 \\
$\sigma_{tot}^{pp} (\sqrt{s}=13\  \textnormal{TeV})$ (mb)& 104.9 & 104.9 & 104.9 & 104.9 & 104.9 \\
$\sigma_{tot}^{\bar{p}p} (\sqrt{s}=13\  \textnormal{TeV})$ (mb) & 105.1 & 105.1 & 105.1 & 105.1 & 105.1 \\
\hline 
\hline \\ [-0.3cm]
 & & \multicolumn{2}{c}{Ensemble $A\oplus T$} & & \\  
$D$ (GeV$^{-2}$) & $0.1A$ & $0.3A$ & $0.5A$ & $0.7A$ & $0.9A$ \\
$\beta_{\Bbb O}(0)$ & 1.09$\pm$0.24 & 0.95$\pm$0.19 & 0.90$\pm$0.18 & 0.86$\pm$0.17 & 0.83$\pm$0.16 \\
$\beta_{\Bbb P}(0)$ & 2.235$\pm$0.023 & 2.257$\pm$0.016 & 2.259$\pm$0.016 & 2.258$\pm$0.016 & 2.258$\pm$0.017 \\
$\nu$ & 504 & 504 & 504 & 504 & 504 \\
$\chi^{2}/\nu$ & 1.11 & 1.12 & 1.11 & 1.10 & 1.09 \\ 
\hline \\ [-0.3cm]
$\rho^{pp} (\sqrt{s}=13\  \textnormal{TeV})$ & 0.112 & 0.112 & 0.111 & 0.111 & 0.110 \\
$\rho^{\bar{p}p} (\sqrt{s}=13\  \textnormal{TeV})$ & 0.119 & 0.118 & 0.119 & 0.119 & 0.120 \\
$\sigma_{tot}^{pp} (\sqrt{s}=13\  \textnormal{TeV})$ (mb)& 104.9 & 104.9 & 104.9 & 104.9 & 104.9 \\
$\sigma_{tot}^{\bar{p}p} (\sqrt{s}=13\  \textnormal{TeV})$ (mb) & 105.1 & 105.1 & 105.1 & 105.1 & 105.1 \\
\end{tabular}
\end{ruledtabular}
\label{tab003}
\end{table*}

\begin{figure*}\label{fig001}
\begin{center}
\includegraphics[height=.99\textheight]{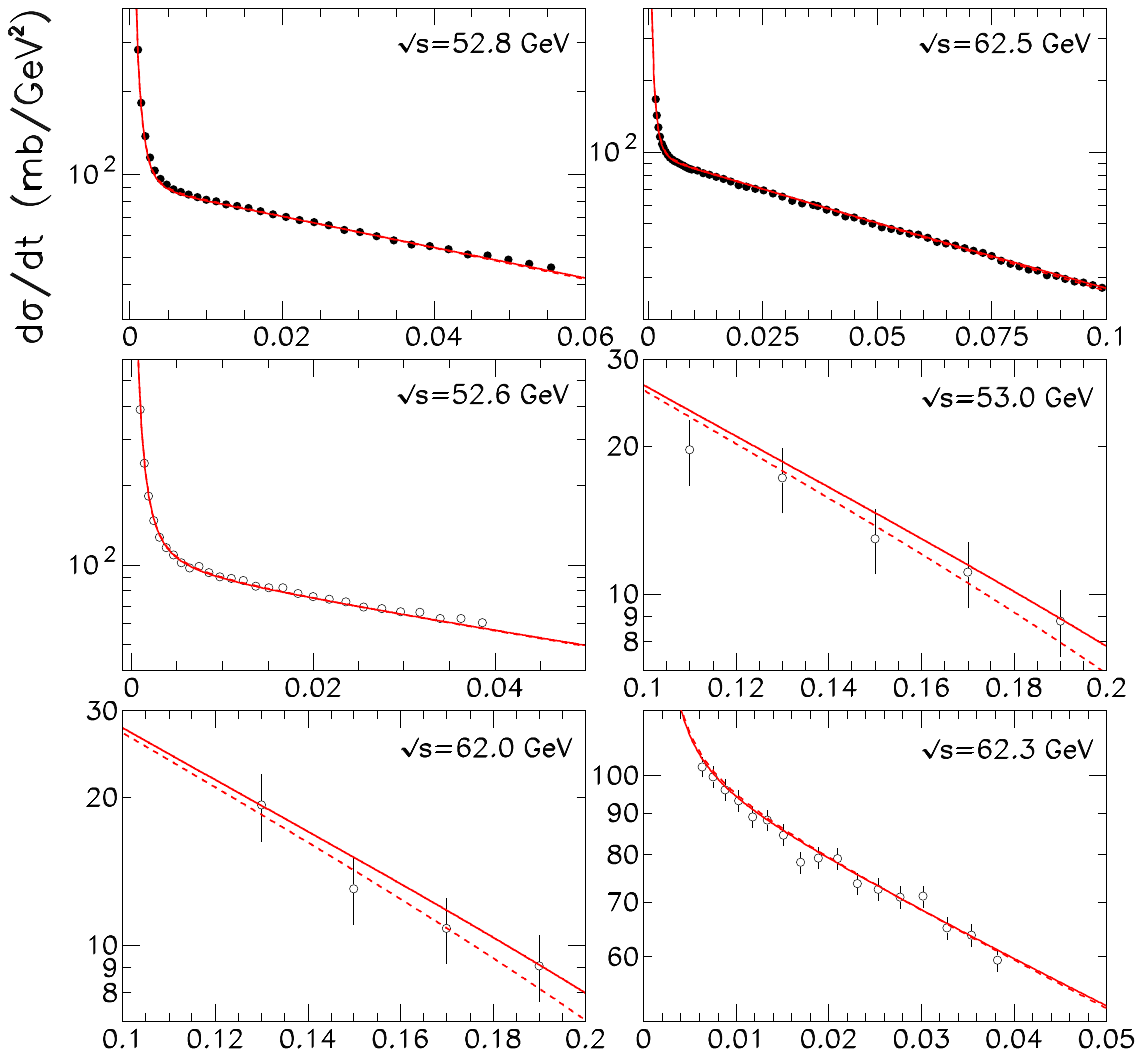}
\caption{Description of the $t$ dependence of the elastic $pp$- and $\bar pp$-cross sections measured at CERN-ISR \cite{exp_amos1,exp_amaldi1,exp_breakstone1}. The dashed and solid curves depict the results obtained using Models I and II, respectively.}
\end{center}
\end{figure*}

\begin{figure*}\label{fig002}
\begin{center}
  \includegraphics[height=.99\textheight]{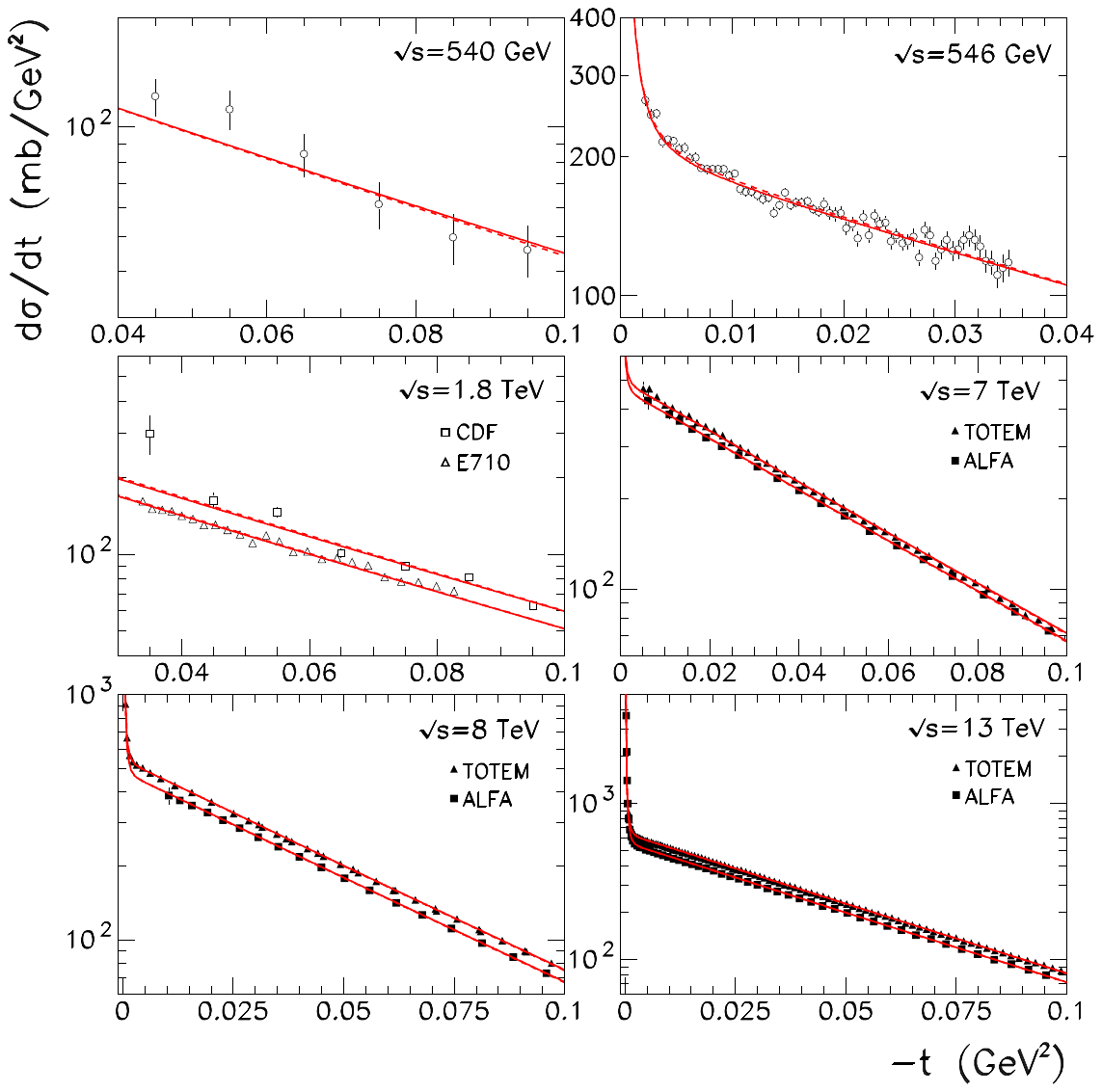}
  \caption{Description of the $t$ dependence of the elastic $pp$- and $\bar pp$-cross sections measured at the $S\bar ppS$ \cite{exp_arnison1,exp_bernard1}, the Tevatron \cite{Tev1,Tev2} and the LHC colliders \cite{LHC1a,LHC1b,LHC1c,Tot-13,Tot-1,LHC2}. The dashed and solid curves depict the results obtained using Models I and II, respectively. The lower curves describe the ATLAS/ALFA (E710) data while the upper curves correspond to the TOTEM (CDF) data; in both cases, the normalization factors $N_i$ are accounted for.}
\end{center}
\end{figure*}

\begin{figure*}
\begin{center}
  \includegraphics[height=.99\textheight]{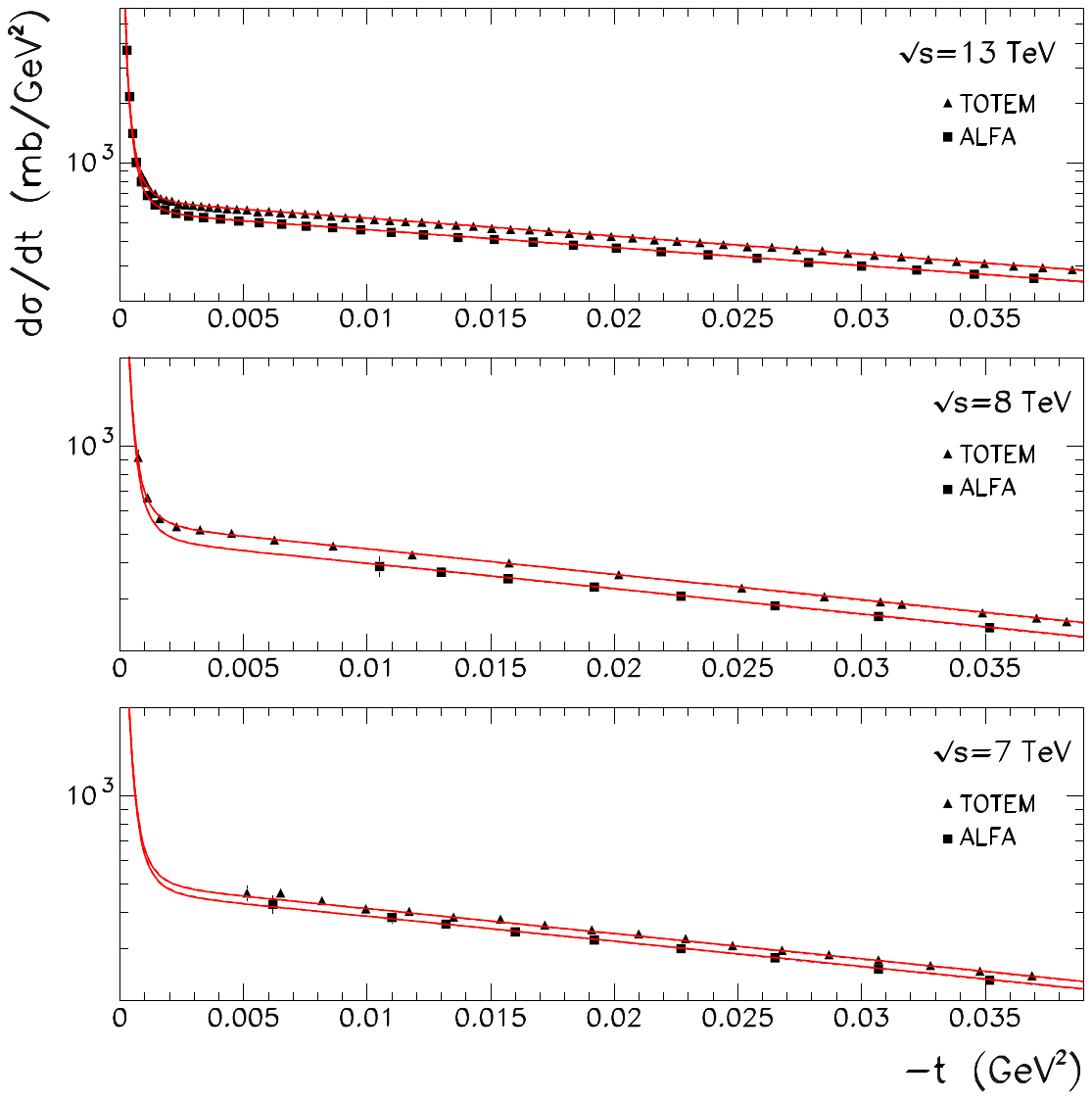}
\caption{The same as Fig.~2 but for the CNI region where the Odderon contribution reveals itself.}
\end{center}
\label{fig003}
\end{figure*}

\begin{figure*}\label{fig004}
\begin{center}
\includegraphics[height=.99\textheight]{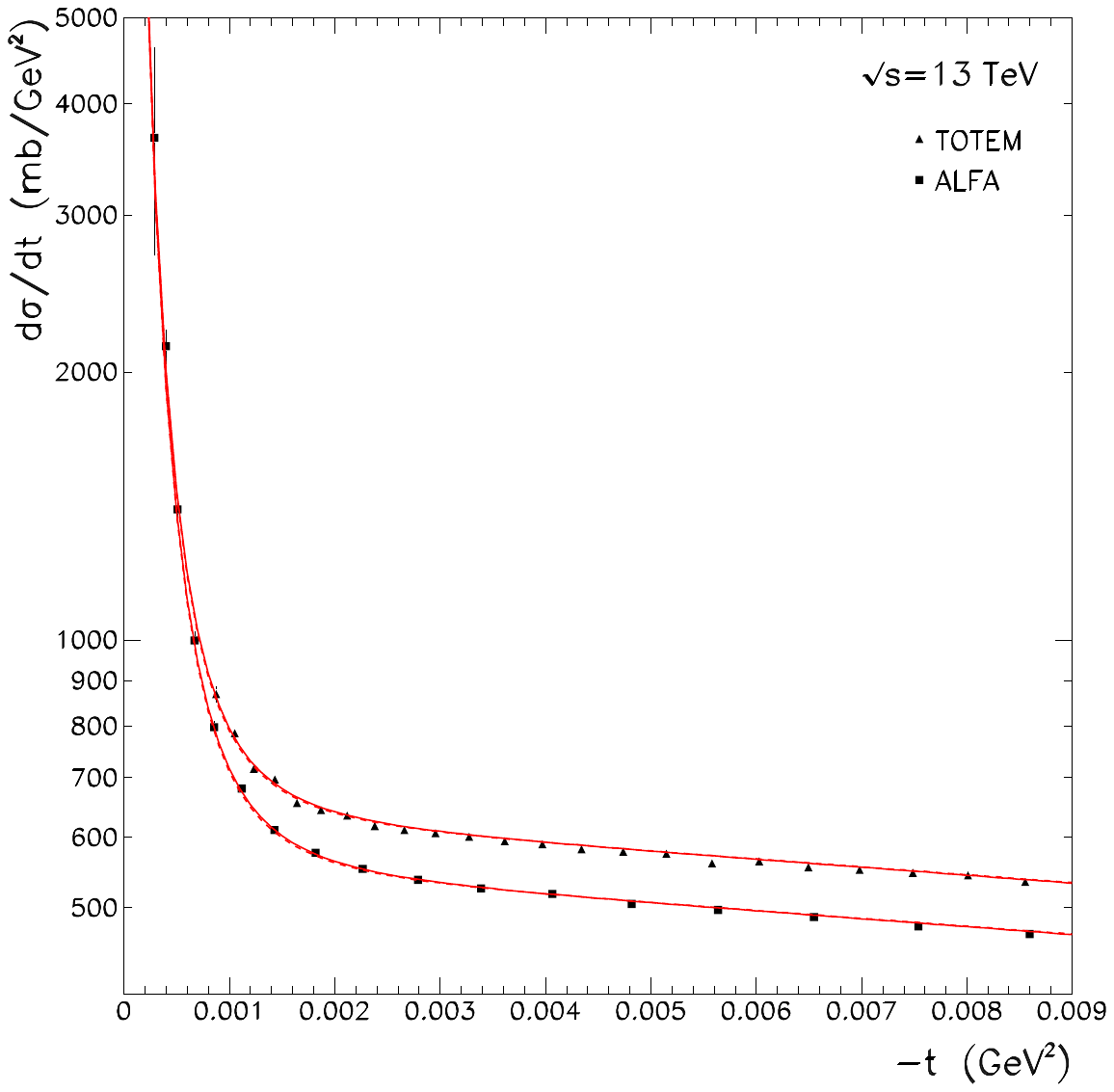}
\caption{The same as Figs.~2 and 3 but in another scale to better see the quality of precise 13 TeV data description.}
\end{center}
\end{figure*}

\begin{figure*}\label{fig005}
\begin{center}
\includegraphics[height=.99\textheight]{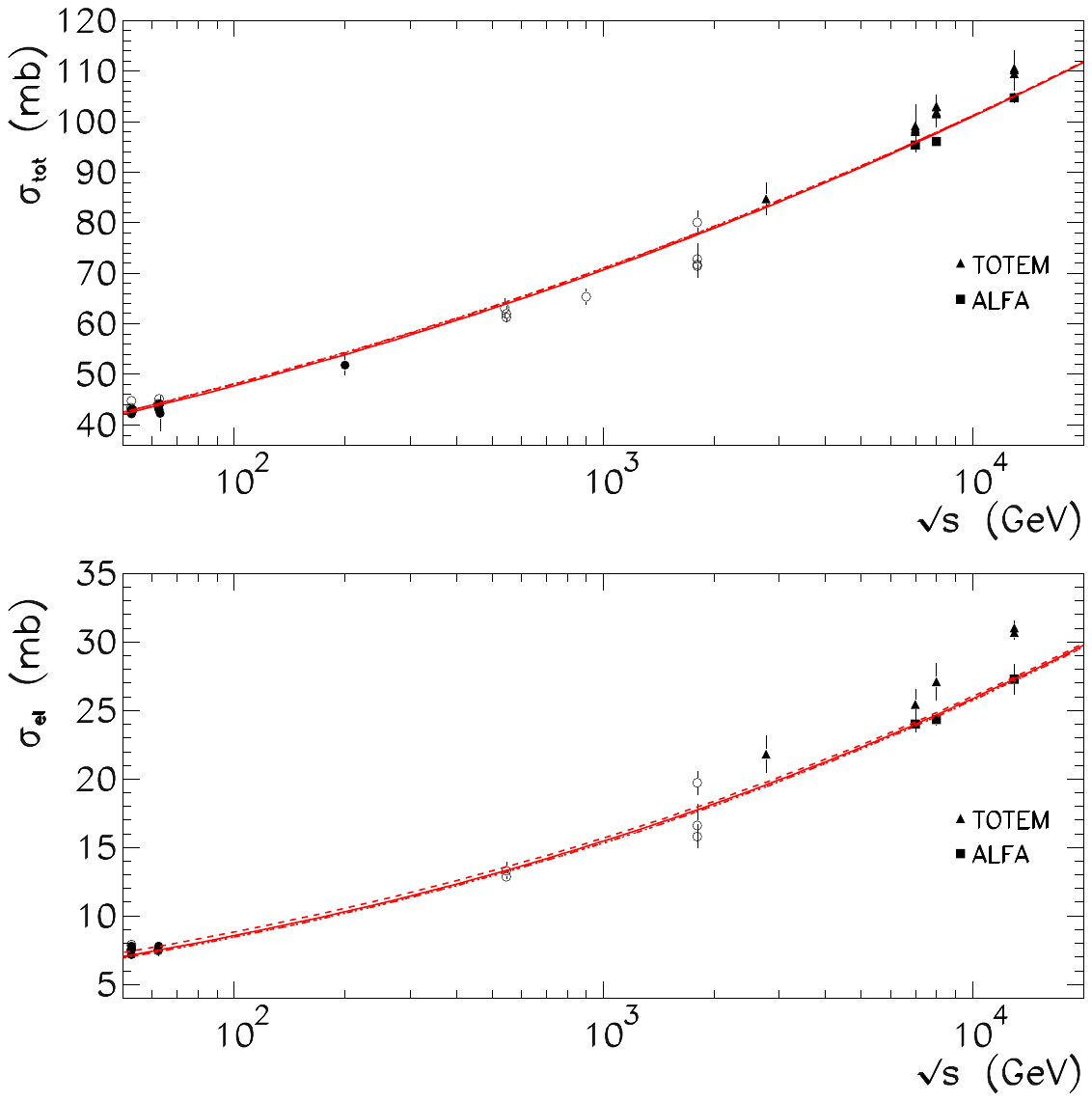}
\caption{Description of the total and elastic $pp$ ($\bullet$, $\blacktriangle$, $\blacksquare$) and $\bar pp$ ($\circ$) cross sections. The data are from \cite{LHC1a,Sigt,LHC2,totem7a,LHC1b}. The dotted and dashed-dotted curves represent the results for $pp$ and $\bar{p}p$ channels, respectively, obtained from the global fit to Ensemble A $\oplus$ T using Model I. These curves are indistinguishable. The solid and dashed curves represent the results for $pp$ and $\bar{p}p$ channels, respectively, obtained from the global fit to Ensemble A $\oplus$ T using Model II.}
\end{center}
\end{figure*}

\begin{figure*}\label{fig006}
\begin{center}
\includegraphics[height=.99\textheight]{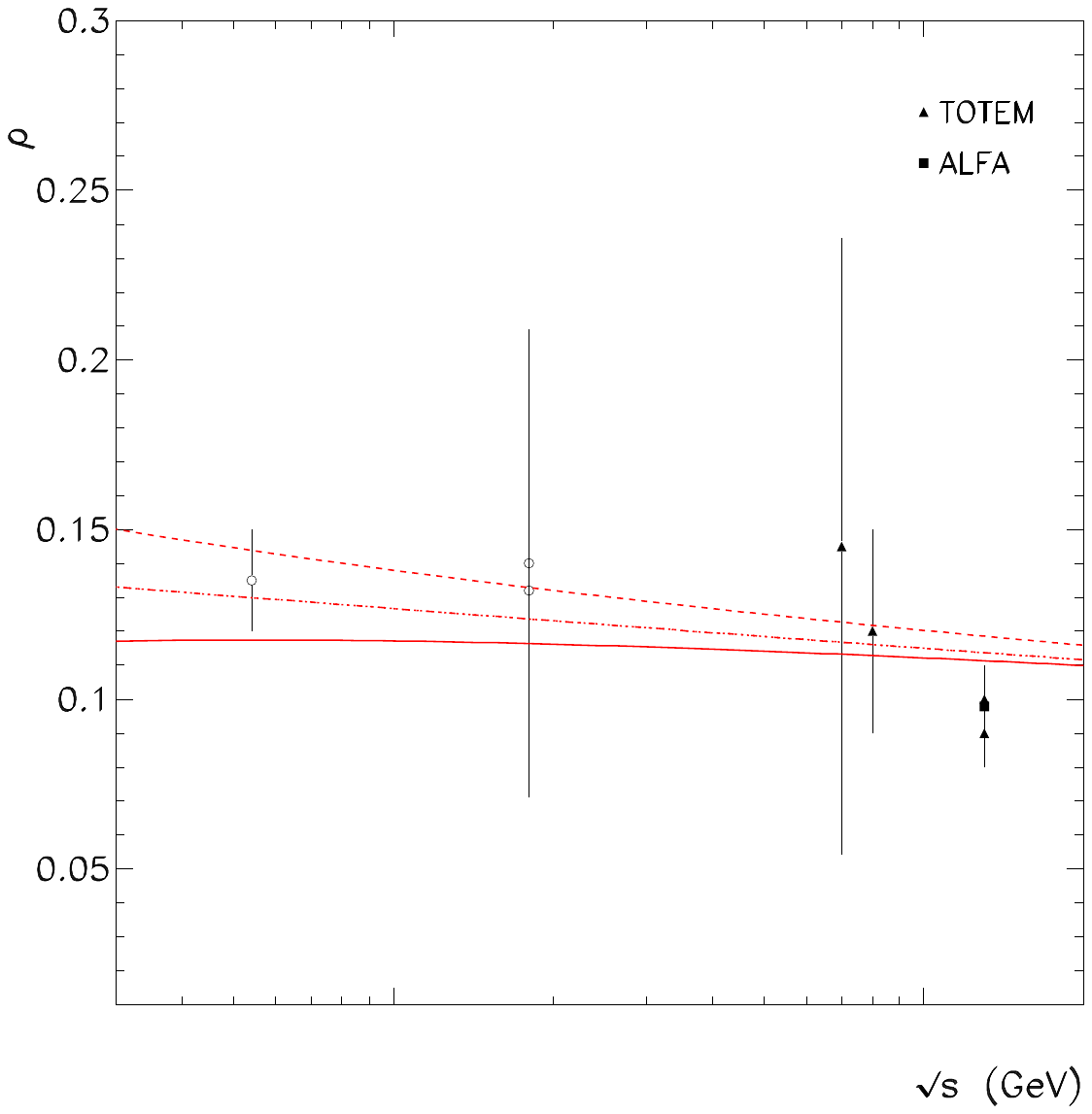}
\caption{$\rho$ parameter for $pp$ ($\blacktriangle$, $\blacksquare$) and $\bar pp$ ($\circ$) elastic amplitude. The data are from \cite{pdg,Tot-2,Tot-1,Tot-13,LHC1c}. The dotted and dashed-dotted curves represent the results for $\rho^{pp}$ and $\rho^{\bar{p}p}$, respectively, obtained from the global fit to Ensemble A $\oplus$ T using the Model I. These curves are indistinguishable. The solid and dashed curves represent the results for $\rho^{pp}$ and $\rho^{\bar{p}p}$, respectively, obtained from the global fit to Ensemble A $\oplus$ T using the Model II.}
\end{center}
\end{figure*}

\begin{figure*}\label{fig007}
\begin{center}
\includegraphics[height=.99\textheight]{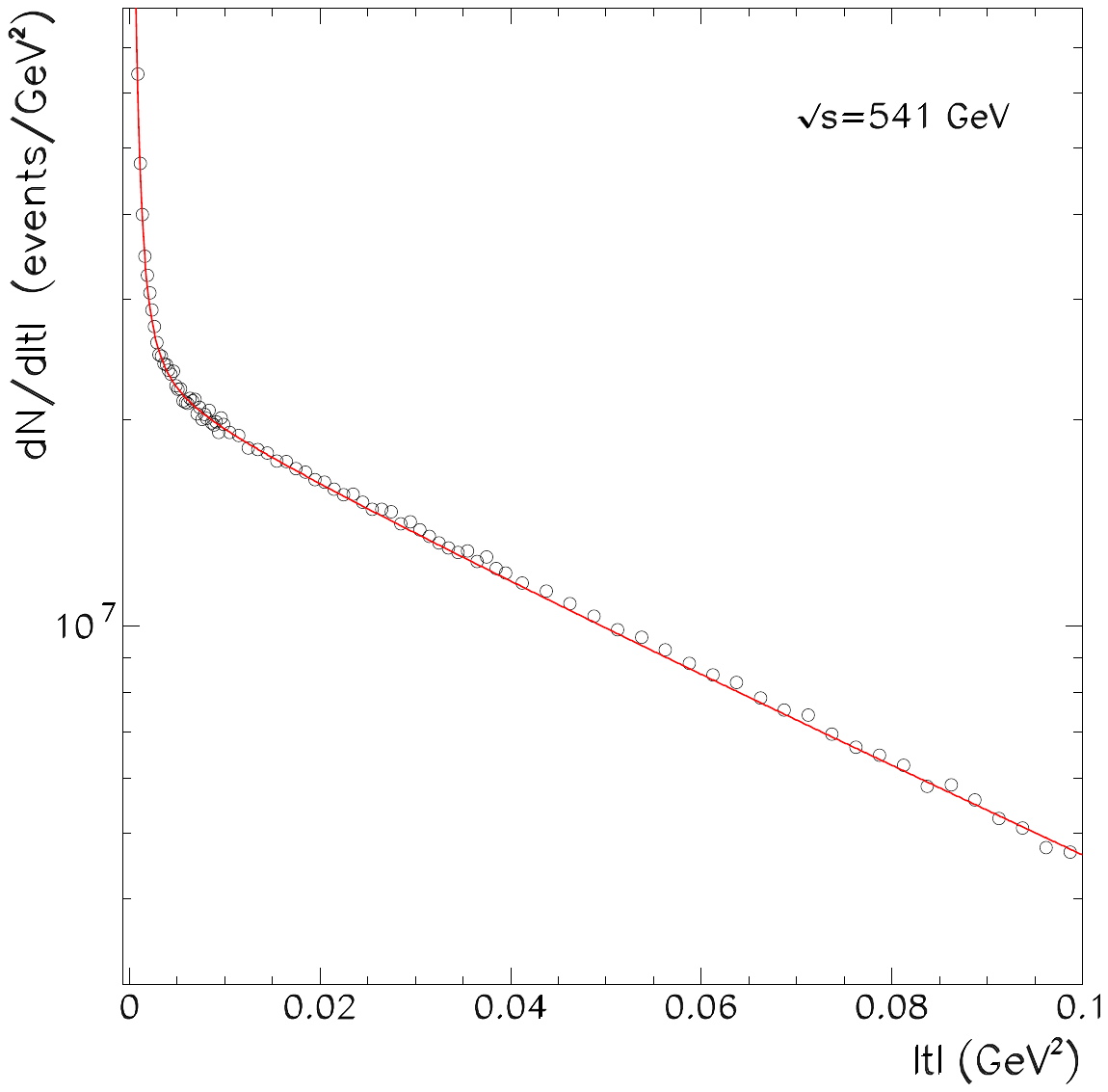}
\caption{Description of the $\bar{p}p$ differential $dN/d|t|$ distribution in the $|t|$-range $|t| \leq 0.1$ GeV$^{2}$.}
\end{center}
\end{figure*}

\begin{figure*}\label{fig008}
\begin{center}
\includegraphics[height=.99\textheight]{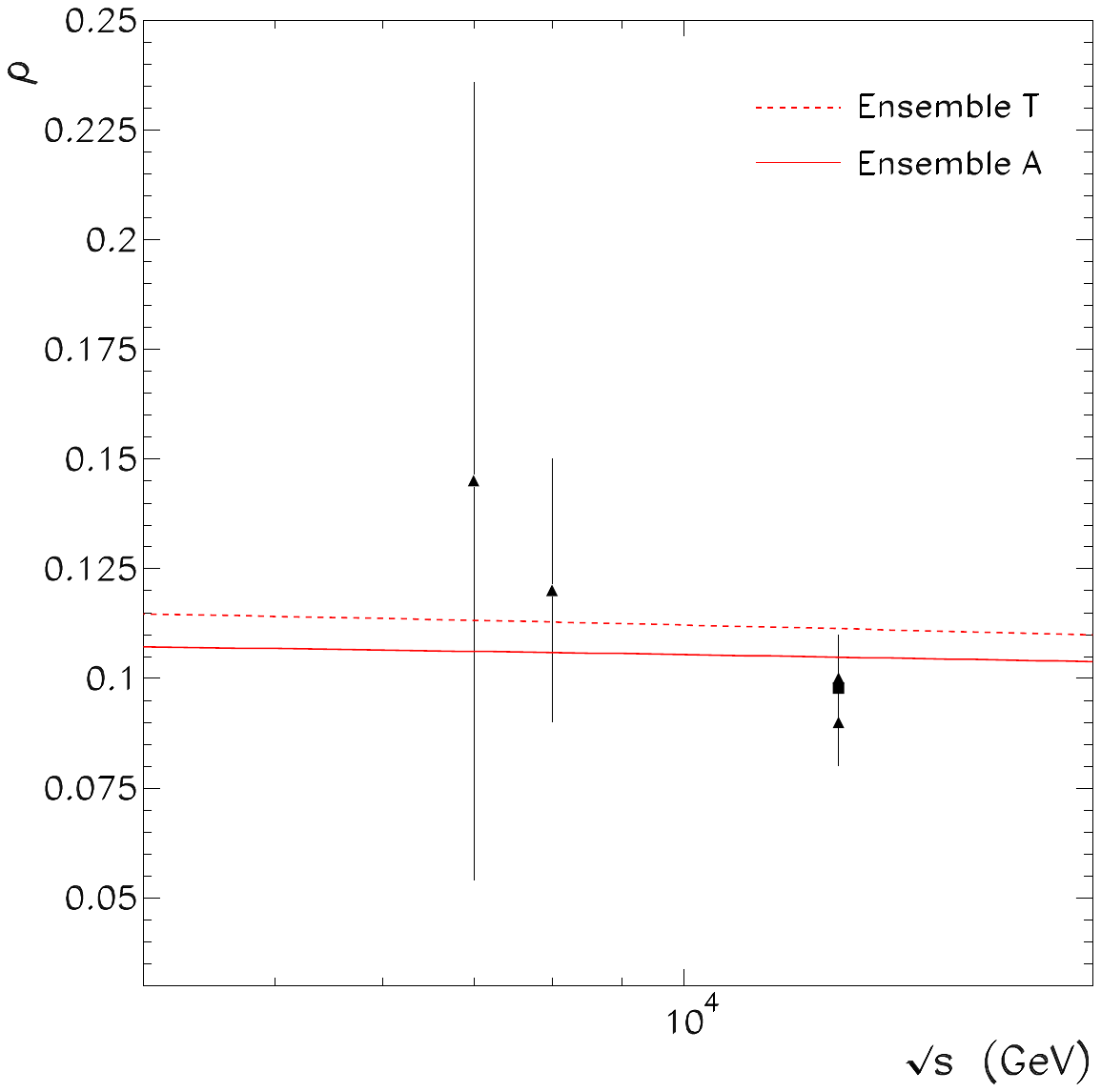}
\caption{Description of $\rho$ parameter for $pp$ elastic amplitude measured by TOTEM ($\blacktriangle$) and ATLAS/ALFA ($\blacksquare$) Collaborations. The data are from \cite{Tot-2,Tot-1,Tot-13,LHC1c}. The dashed (solid) curve represents the predicted $\rho^{pp}$ from the global fit using Ensemble T (Ensemble A).}
\end{center}
\end{figure*}

\end{document}